# Machine vs Machine: Using AI to Tackle Generative AI Threats in Assessment

Mohammad Saleh Torkestani
*faculty of environment science and economy*
*University of Exeter*
Exeter, United Kingdom
m.torkestani@exeter.ac.uk

Taha Mansouri
*School of Science, Engineering and Environment*
*University of Salford*
Manchester, United Kingdom
t.mansouri@salford.ac.uk

**Abstract**

This paper presents a theoretical framework for addressing the challenges posed by generative artificial intelligence (AI) in higher education assessment through a machine-versus-machine approach. Large language models like GPT-4, Claude, and Llama increasingly demonstrate the ability to produce sophisticated academic content, traditional assessment methods face an existential threat, with surveys indicating 74-92% of students experimenting with these tools for academic purposes. Current responses, ranging from detection software to manual assessment redesign, show significant limitations: detection tools demonstrate bias against non-native English writers and can be easily circumvented, while manual frameworks rely heavily on subjective judgment and assume static AI capabilities. This paper introduces a dual strategy paradigm combining static analysis and dynamic testing to create a comprehensive theoretical framework for assessment vulnerability evaluation. The static analysis component comprises eight theoretically justified elements: specificity and contextualization, temporal relevance, process visibility requirements, personalization elements, resource accessibility, multimodal integration, ethical reasoning requirements, and collaborative elements. Each element addresses specific limitations in generative AI capabilities, creating barriers that distinguish authentic human learning from AI-generated simulation. The dynamic testing component provides a complementary approach through simulation-based vulnerability assessment, addressing limitations in pattern-based analysis. The paper presents a theoretical framework for vulnerability scoring, including the conceptual basis for quantitative assessment, weighting frameworks, and threshold determination theory. We discuss implications for assessment design, limitations of the dual strategy approach, and directions for future development. This machine-versus-machine paradigm offers a robust foundation for maintaining assessment integrity while enhancing pedagogical effectiveness in an AI-augmented educational landscape.

# 1. Introduction

In recent years, higher education has witnessed an unprecedented technological revolution with the emergence of generative artificial intelligence (AI) tools. These sophisticated systems, powered by large language models (LLMs) such as GPT-4, Claude, and Llama, have transformed how students approach learning, research, and assessment tasks (Yusuf et al., 2024). While these technologies offer remarkable opportunities for enhancing education through personalized learning experiences and administrative efficiencies, they simultaneously present profound challenges to traditional assessment paradigms and academic integrity frameworks (Bittle & El-Gayar, 2025). The ability of generative AI to produce coherent, contextually relevant, and seemingly original content has created what many educators describe as an existential crisis for conventional assessment methods (Flaherty, 2025).

The scale of this challenge cannot be overstated. Recent surveys indicate that between 74% and 92% of students have experimented with generative AI tools for academic purposes, with a significant proportion using these technologies specifically for assessment tasks (Higher Education Policy Institute, 2025; Weale, 2025). Chief technology officers across institutions report that artificial intelligence has proven to be a moderate or significant risk to academic integrity, with three in four expressing serious concerns about its impact (Flaherty, 2025). Traditional assessments, often based on static case studies, standardized questions, or dated examples, have become particularly vulnerable to AI-generated responses that can produce formulaic yet convincing answers that evade conventional plagiarism detection systems (Sadasivan et al., 2023).

The limitations of current approaches to addressing this challenge are becoming increasingly apparent. Manual detection methods struggle to keep pace with rapidly evolving AI capabilities, while automated detection tools face significant accuracy and equity concerns. Jiang et al. (2024) demonstrate that standard detectors mis-label genuine essays by advanced non-native English writers at five times the rate recorded for native writers, raising serious questions about fairness in implementation. Furthermore, simple text transformations can dramatically reduce detection rates, with Sadasivan et al. (2023) showing that editing only a small fraction of AI-generated output can collapse detector recall to below 10%. Regulatory bodies have consequently cautioned against high-stakes reliance on detector scores alone (Ofqual, 2024). In response to these challenges, educators have developed various frameworks for redesigning assessments to be more resilient to AI-generated content. Approaches such as the MAGE cycle (Map, Attempt, Grade, Evaluate) proposed by Zaphir et al. (2024) and the AI Assessment Scale (AIAS) by Perkins et al. (2024) offer valuable heuristics. However, these frameworks share two critical limitations: they rely heavily on subjective judgment, and they assume a static snapshot of AI capability rather than accounting for its rapid evolution. Moreover, the time constraints facing educators mean that only a handful of assessment briefs can be manually audited before deadlines, leaving program-level coverage elusive.

This paper proposes a fundamentally different approach: using AI to tackle AI, a machine-versus-machine strategy that leverages a dual methodology combining static analysis and dynamic testing. This dual strategy draws on principles from cybersecurity, educational theory, and assessment design to create a comprehensive framework for evaluating and enhancing assessment resilience to generative AI. Unlike existing approaches that rely primarily on either pattern recognition or post-submission detection, the dual strategy provides a proactive, theoretically grounded methodology for identifying and addressing vulnerabilities before assessments are distributed to students.

At the heart of this approach is a theoretical framework for static analysis comprising eight distinct elements: specificity and contextualization, temporal relevance, process visibility requirements, personalization elements, resource accessibility, multimodal integration, ethical reasoning requirements, and collaborative elements. Each element is grounded in established educational theories and represents a distinct dimension along which assessment vulnerability can be evaluated. When combined with dynamic testing, the simulation of AI-generated responses to assessment tasks, these elements provide a comprehensive basis for understanding and addressing the challenges posed by generative AI. The justification for this dual strategy draws on multiple disciplinary perspectives. From cybersecurity comes the concept of red-teaming, proactively testing systems for vulnerabilities before they can be exploited. From educational theory comes the emphasis on constructive alignment (Biggs & Tang, 2022), authentic assessment (Honea et al., 2017), and feedback literacy (Carless & Boud, 2018). From assessment design comes the focus on validity, reliability, and fairness in measuring student learning. By integrating these perspectives, the dual strategy offers a theoretically robust approach to maintain assessment integrity in an era of ubiquitous generative AI.

This paper makes several contributions to the ongoing discourse on AI in higher education. First, it provides a conceptual framework for understanding how machine-versus-machine approaches can maintain assessment integrity. Second, it offers a detailed justification for the eight elements of static analysis, explaining how each addresses specific vulnerabilities in assessment design. Third, it presents a theoretical model for integrating static and dynamic approaches, demonstrating how their complementary strengths can overcome their respective limitations. Finally, it discusses broader implications for assessment theory, educational philosophy, and the future of learning in an AI-augmented world. The remainder of this paper is structured as follows: Section 2 reviews the literature on generative AI in higher education, challenges to assessment integrity, frameworks for AI-resilient assessment, and red-teaming as a construct. Section 3 presents the foundations of the dual strategy approach, including the conceptual basis for static analysis, dynamic testing, and their synergistic integration. Section 4 provides a detailed justification for the eight elements of static analysis, explaining how each addresses specific vulnerabilities in assessment design. Section 5 presents a theoretical framework for vulnerability scoring, including the conceptual basis for quantitative assessment, weighting frameworks, and threshold determination. Section 6 discusses implications, limitations, and future directions. Finally, Section 7 concludes with a summary of contributions and broader implications for assessment in higher education.

## 2. Theoretical Foundations

### 2.1. Conceptual Basis for Static Analysis

Static analysis in the context of assessment vulnerability represents a theoretically grounded approach to identifying potential weaknesses in assessment design before implementation. Drawing from software engineering principles, where static code analysis examines program structure without execution (Brundage et al., 2020), educational static analysis evaluates assessment briefs against established patterns of vulnerability without actually attempting to complete the assessment. This approach is fundamentally pattern-recognition based, examining structural and content characteristics that correlate with susceptibility to generative AI exploitation. The justification for static analysis rests on several foundational premises. First, certain assessment characteristics consistently predict vulnerability to AI-generated responses across different contexts and disciplines. These patterns emerge from the inherent limitations and capabilities of large language models, which excel at generating responses to generic, fact-based, or historically well-documented prompts but struggle with highly specific, novel, or multimodal tasks (Floridi & Chiriatti, 2020). By identifying these patterns through systematic analysis, educators can proactively modify assessment designs to reduce vulnerability. Second, static analysis aligns with established assessment design principles, particularly constructive alignment theory (Biggs & Tang, 2022). Constructive alignment emphasizes the importance of coherence between learning outcomes, teaching activities, and assessment tasks. Static analysis extends this framework by considering an additional dimension: resilience to generative AI. This integration ensures that modifications to enhance AI resilience do not compromise the fundamental pedagogical integrity of the assessment, maintaining alignment with intended learning outcomes while reducing vulnerability. Third, static analysis offers a scalable approach to assessment evaluation. Unlike methods requiring individual judgment for each assessment, pattern-based analysis can be systematized and applied consistently across large numbers of assessments. This scalability addresses a critical limitation of manual frameworks, which typically allow educators to audit only a handful of assessments before implementation deadlines (Zaphir et al., 2024). By enabling program-level coverage, static analysis provides a more comprehensive approach to maintaining assessment integrity across entire curricula.

Despite these strengths, static analysis in isolation faces significant limitations. Most notably, it relies on historical patterns of vulnerability, which may not fully capture the rapidly evolving capabilities of generative AI. As Kaplan et al. (2020) demonstrate through scaling laws for neural language models, AI capabilities improve predictably with increases in computational resources and training data, potentially rendering previously effective patterns obsolete. This limitation suggests that static analysis alone provides an incomplete picture of assessment vulnerability. Additionally, static analysis cannot account for the creative ways in which students might leverage AI tools to address specific assessment tasks. The gap between vulnerability patterns and actual AI performance on particular assessments may be substantial, especially for assessments with unique or innovative structures. This limitation echoes concerns in cybersecurity, where static vulnerability scanning often misses context-specific exploits that emerge only during active testing (Brundage et al., 2020).

Finally, static analysis typically focuses on textual elements, potentially overlooking vulnerabilities in multimodal assessments or those requiring specialized knowledge representation. As generative AI increasingly encompasses multimodal capabilities, including image, audio, and code generation, the patterns of vulnerability become more complex and less amenable to purely text-based static analysis. This limitation points to the need for complementary approaches that can address the dynamic, evolving nature of AI capabilities.

## 2.2. Theoretical Basis for Dynamic Testing

Dynamic testing represents a fundamentally different approach to assessment vulnerability evaluation. Rather than analyzing patterns in assessment design, dynamic testing actively simulates how generative AI might respond to the assessment task, providing empirical evidence of vulnerability through actual performance. This approach draws conceptual foundations from adversarial thinking in cybersecurity, where systems are actively probed for weaknesses rather than theoretically evaluated (Brundage et al., 2020).

The justification for dynamic testing begins with the recognition that generative AI capabilities are not fully predictable from static characteristics alone. As Lewis et al. (2020) demonstrate, large language models exhibit emergent capabilities that may not be apparent from their architecture or training methodology. These emergent properties mean that AI performance on specific tasks can sometimes surprise even their developers, making predictions insufficient for comprehensive vulnerability assessment. Dynamic testing addresses this limitation by directly observing AI performance on the assessment task. Second, dynamic testing aligns with the educational concept of authentic assessment (Honea et al., 2017), which emphasizes tasks that mirror real-world challenges and applications. By simulating actual student use of generative AI, dynamic testing provides an authentic evaluation of how the assessment might perform under real conditions. This authenticity enhances the validity of vulnerability assessment, ensuring that it reflects genuine rather than theoretical risks. Third, dynamic testing captures the iterative, feedback-driven nature of AI use in educational contexts. Students rarely use generative AI in a single, isolated attempt; instead, they typically engage in multiple rounds of prompting, refinement, and integration (Kumar & Upadhyay, 2024). Dynamic testing can simulate this iterative process, providing insight into how persistent or sophisticated AI use might overcome initial barriers in the assessment design. This iterative perspective offers a more nuanced understanding of vulnerability than static analysis alone.

However, dynamic testing also faces significant limitations when used in isolation. Most notably, it provides specific rather than generalizable insights, demonstrating vulnerability for particular AI models and prompting strategies but not necessarily identifying the underlying patterns that create vulnerability. This specificity limits the transferability of findings across different assessment contexts or as AI capabilities evolve, potentially requiring repeated testing for each new assessment or AI advancement. Additionally, dynamic testing is resource-intensive compared to static analysis, requiring actual execution of AI models against assessment tasks. This resource requirement creates practical constraints on the number of assessments that can be evaluated and the thoroughness of the evaluation. As Chaudhry et al. (2022) note, resource constraints often force trade-offs between breadth and depth in educational AI applications, potentially limiting the comprehensiveness of dynamic testing approaches. Finally, dynamic testing may not fully capture the human-AI collaboration that characterizes actual student use of generative tools. Students bring domain knowledge, contextual understanding, and strategic thinking to their AI interactions, potentially enabling more sophisticated responses than automated testing would predict. This limitation suggests that dynamic testing, while valuable, provides an incomplete picture of real-world assessment vulnerability.

### 2.3. Synergistic Integration: The Dual Strategy Paradigm

The limitations of both static analysis and dynamic testing, when used in isolation, point to the need for an integrated approach that leverages their complementary strengths. The dual strategy paradigm represents such an integration, combining pattern-based vulnerability analysis with simulation-based performance testing to provide a comprehensive theoretical framework for assessment evaluation. The justification for this integration rests on several key principles. First, the complementary nature of the two approaches addresses their respective limitations. Static analysis provides scalable, pattern-based evaluation but may miss emerging or context-specific vulnerabilities; dynamic testing captures specific performance but may not identify underlying patterns. By combining these approaches, the dual strategy paradigm offers both breadth and depth in vulnerability assessment, enabling both program-level coverage and detailed insight into particular assessments. Second, the dual strategy aligns with established principles in risk assessment theory, particularly the distinction between vulnerability identification and exploitation verification. In cybersecurity frameworks, potential vulnerabilities are first identified through pattern recognition, then verified through actual exploitation attempts (Brundage et al., 2020). This two-stage approach reduces false positives (identifying vulnerabilities that cannot actually be exploited) and false negatives (missing vulnerabilities that do not match recognized patterns), enhancing the overall accuracy of assessment. Third, the integration of static and dynamic approaches creates a feedback loop that enhances the robustness of both components. Patterns identified through static analysis can guide more targeted dynamic testing, while the results of dynamic testing can refine and update the patterns used in static analysis. This iterative refinement addresses the challenge of rapidly evolving AI capabilities, ensuring that vulnerability assessment remains current and effective even as new models and techniques emerge.

The conceptual framework for this integration involves several key components. First, static analysis provides an initial vulnerability assessment based on established patterns, identifying potential weaknesses in the assessment design. This preliminary evaluation guides the focus of subsequent dynamic testing, ensuring efficient use of resources by concentrating on areas of greatest vulnerability. Second, dynamic testing verifies and refines the static analysis, providing empirical evidence of actual rather than vulnerability. This verification step helps distinguish between genuine vulnerabilities that require intervention and concerns that may not manifest in practice. The results of dynamic testing also provide specific examples of how generative AI might approach the assessment task, offering concrete insights for remediation. Third, the results of both approaches are integrated into a comprehensive vulnerability assessment that considers both pattern-based and performance-based evidence. This integration provides a more nuanced understanding of assessment vulnerability than either approach alone, capturing both generalizable patterns and specific performance characteristics. The combined assessment forms the basis for targeted remediation strategies that address verified vulnerabilities while maintaining the pedagogical integrity of the assessment.

The dual strategy paradigm represents a significant advancement over existing approaches to assessment vulnerability. By integrating static and dynamic methodologies, it offers a more comprehensive, accurate, and adaptable framework for evaluating and enhancing assessment resilience to generative AI. This integration addresses the fundamental challenge of assessment in an AI-augmented educational landscape: maintaining validity and integrity while acknowledging the rapidly evolving capabilities of generative technologies.

The diagram below illustrates the dual strategy workflow, which begins with static analysis to detect structural vulnerabilities in assessment design and continues with dynamic testing to empirically validate risks. The process is iterative, enabling revisions based on both theoretical patterns and AI response behaviours.

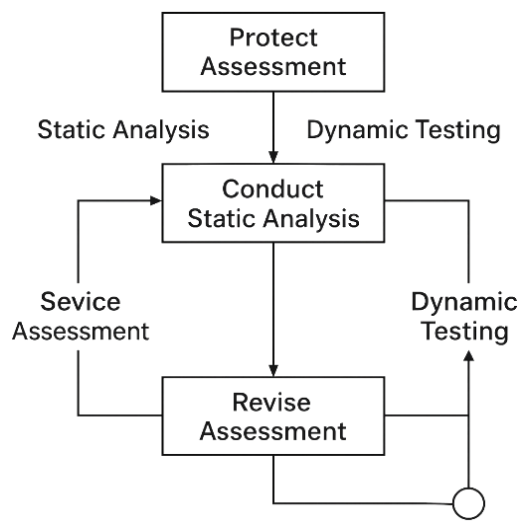

Figure 1 - The dual strategy workflow

## 3. The Eight Elements of Static Analysis: Theoretical Justification

To proactively evaluate assessment vulnerability in the face of generative AI, this section introduces eight theoretically grounded elements that form the core of the static analysis framework. Each element represents a dimension along which an assessment can either resist or succumb to AI-generated content. These elements are designed not only to exploit known limitations in large language models but also to align with established educational principles such as constructive alignment, authentic assessment, and metacognitive development. The table below summarizes each element, its key operational dimensions, and the specific AI vulnerability it is intended to target.

Table 1 - Eight Elements of Static Analysis

|   | Element | Key Dimensions | AI Vulnerability Targeted |
|---|---------|----------------|---------------------------|
| 1 | Specificity & Contextualization | Topical, Contextual, Analytical Specificity | Pattern-based generalisation |
| 2 | Temporal Relevance | Reference Recency, Event Recency, Analytical Recency | Training data cutoff limitations |
| 3 | Process Visibility Requirements | Developmental, Justificatory, Reflective Visibility | Lack of iterative thought simulation |
| 4 | Personalization Elements | Experiential, Reflective, Applicative Personalization | No personal experience or reflection |
| 5 | Resource Accessibility | Exclusivity, Specificity, Format Diversity | Restricted to training corpus or prompt input |
| 6 | Multimodal Integration | Cross-modal, Synthesis, Representational Translation | Limited cross-modal synthesis |
| 7 | Ethical Reasoning Requirements | Identification, Analysis, Resolution of Dilemmas | Lack of genuine moral agency |
| 8 | Collaborative Elements | Interactive, Integrative, Negotiated Collaboration | No real-time interaction or negotiation |

## 3.1. Specificity and Contextualization

Specificity and contextualization represent foundational elements in the theoretical framework for static analysis of assessment vulnerability. This element focuses on the degree to which an assessment requires responses that are unique, highly specific, and contextually situated rather than generic or broadly applicable. The theoretical basis for including specificity as a defense against AI-generated responses stems from the fundamental architecture and training methodology of large language models.

Large language models operate by predicting the most probable next token based on patterns observed in their training data (Kaplan et al., 2020). This probabilistic approach excels at generating content that follows common patterns and addresses general topics but struggles with highly specific or unique contexts that appear infrequently in training data. As Floridi and Chiriatti (2020) observe, these models lack genuine understanding of novel or specific contexts, instead relying on statistical correlations derived from similar contexts encountered during training. This limitation creates a theoretical vulnerability in assessments that rely on generic, widely discussed topics or standard analytical frameworks. The justification for specificity as a defense mechanism aligns closely with constructive alignment principles (Biggs & Tang, 2022). Constructive alignment emphasizes that assessment tasks should directly measure the specific learning outcomes of a particular educational experience. By increasing specificity and contextualization, assessments not only become more resistant to AI-generated responses but also more closely aligned with the unique learning journey of students in a specific course or program. This dual benefit enhances both assessment security and pedagogical validity.

Specificity can be conceptualized along several dimensions. First, topical specificity refers to the uniqueness of the subject matter addressed in the assessment. Assessments focusing on niche topics, recent developments, or specialized applications present greater challenges for generative AI than those addressing widely discussed subjects with extensive documentation in training data. Second, contextual specificity involves embedding the assessment in particular circumstances, cases, or scenarios that require familiarity with details not widely available.

Third, analytical specificity demands unique frameworks or approaches rather than standard methodologies that appear frequently in academic and professional literature.

Conceptual measurement approaches for specificity might include evaluating the uniqueness of key terms, analyzing the prevalence of assessment-specific concepts in general corpora, and assessing the degree to which successful completion requires engagement with course-specific materials rather than general knowledge. These measurement approaches provide a basis for quantifying specificity as a dimension of assessment vulnerability. However, the framework must also acknowledge potential tensions between specificity and other educational values. Excessive specificity might compromise the transferability of learning or create artificial constraints that do not reflect authentic professional practice. The challenge lies in balancing specificity as a defense mechanism with the broader educational goal of developing transferable knowledge and skills. This balance requires thoughtful integration of specificity with other elements of the static analysis framework.

### 3.2. Temporal Relevance

Temporal relevance emerges as a critical element in the theoretical framework for static analysis, focusing on the recency and time-bound nature of assessment content. This element addresses a fundamental limitation of large language models: their training data has a cutoff date, beyond which they lack direct knowledge of events, developments, or publications (OpenAI, 2023). This temporal constraint creates a theoretical basis for enhancing assessment resilience through the strategic incorporation of recent content. The justification for temporal relevance as a vulnerability reducer stems from the architectural constraints of generative AI systems. Unlike humans, who continuously update their knowledge through ongoing experience and learning, large language models operate with fixed training datasets that represent knowledge up to a specific point in time. While some models incorporate limited retrieval capabilities to access more recent information, their core parameters remain tied to their training cutoff. This limitation creates a vulnerability in assessments that relies exclusively on historical or well-established content but offers a potential defense mechanism for those incorporating very recent developments.

From an educational theory perspective, temporal relevance aligns with the principle of currency in professional education. Honea et al. (2017) emphasize that authentic assessment should reflect current professional practices and contemporary challenges rather than historical scenarios that may no longer represent the field. By incorporating recent developments, assessments not only become more resistant to AI-generated responses but also better prepare students for the current state of their chosen profession. This alignment between security and pedagogical value strengthens the theoretical case for temporal relevance as an assessment element. Conceptual implementation considerations for temporal relevance include several dimensions. First, reference recency involves citing or requiring engagement with sources published after the training cutoff dates of common AI models. Second, event recency incorporates recent events, developments, or case studies that post-date AI training data. Third, analytical recency requires application of emerging frameworks or methodologies that have gained prominence after AI training cutoffs. These dimensions provide a theoretical basis for operationalizing temporal relevance in assessment design.

However, the theoretical framework must also acknowledge limitations and challenges associated with temporal relevance. Most notably, the effectiveness of this element diminishes as AI models are updated with more recent training data, creating a moving target for assessment designers. Additionally, excessive emphasis on recency might undermine the assessment of foundational knowledge or historical understanding that remains valuable despite not being current. These limitations suggest that temporal relevance, while theoretically sound, must be integrated with other elements rather than relied upon exclusively. Furthermore, justification must consider equity implications of heavily emphasizing temporal relevance. Students with limited access to current resources or those in regions with delayed access to recent publications might be disadvantaged by assessments that place excessive weight on very recent developments. This consideration highlights the importance of balancing temporal relevance with accessibility and fairness in the overall framework.

### 3.3. Process Visibility Requirements

Process visibility requirements constitute a theoretically significant element in the static analysis framework, focusing on the degree to which assessments demand explicit demonstration of thinking processes rather than merely final products or conclusions. This element addresses a key limitation of generative AI: while large language models excel at producing polished outputs, they typically do not authentically capture the messy, iterative nature of human thinking and problem-solving (Zaphir et al., 2024). The theoretical basis for emphasizing process over product stems from fundamental differences between human and AI cognition. Human learning involves false starts, reconsiderations, and conceptual evolution that reflect genuine engagement with material. In contrast, generative AI produces outputs through a fundamentally different process of token prediction based on statistical patterns (Floridi & Chiriatti, 2020). By requiring visible evidence of the thinking process, including preliminary ideas, revisions, and reflections on decision points, assessments can theoretically distinguish between authentic human learning and AI-generated content that lacks this developmental trajectory.

This element aligns closely with metacognitive learning theories, which emphasize the importance of thinking about thinking in educational development. As Carless and Boud (2018) argue, developing students' ability to monitor, evaluate, and regulate their own cognitive processes represents a crucial educational outcome. Process visibility requirements not only enhance assessment security but also promote these valuable metacognitive skills, creating alignment between AI resilience and pedagogical value.

Conceptual measurement approaches for process visibility might include evaluating the degree to which assessments require documentation of preliminary research, explicit justification of methodological choices, reflection on alternative approaches considered, or annotation of thought processes. These requirements create barriers for generative AI, which can produce polished final products but struggle to authentically simulate the developmental trajectory of human thinking.

However, the theoretical framework must acknowledge potential tensions between process visibility and other educational values. Excessive focus on process documentation might create artificial constraints that do not reflect authentic professional practice in fields where final products are primarily valued. Additionally, process requirements might disadvantage students with different thinking styles or those from cultural backgrounds where explicit process articulation is less emphasized. These considerations highlight the importance of thoughtfully implementing process visibility requirements within the broader educational context. process visibility can be conceptualized along several dimensions. First, developmental visibility involves documenting how ideas or approaches evolved over time. Second, justification requires explicit reasoning for choices made during the assessment process. Third, reflective visibility demands metacognitive consideration of strengths, limitations, and alternatives. These dimensions provide a theoretical basis for operationalizing process visibility in assessment design while maintaining alignment with educational objectives.

### 3.4. Personalization Elements

Personalization elements represent a theoretically distinct component of the static analysis framework, focusing on the degree to which assessments require integration of students' unique experiences, perspectives, or circumstances. This element addresses a fundamental limitation of generative AI: while large language models can produce generic responses or even simulate perspectives, they cannot authentically incorporate truly personal experiences or genuine individual reflection (Kshetri et al., 2024). The justification for personalization as an AI challenge stems from the nature of large language models as statistical systems trained on public data. These models lack genuine personal experience or individual consciousness, instead generating content based on patterns observed across their training corpus. When assessments require integration of specific personal experiences, reflections on individual learning journeys, or application to unique circumstances, they create barriers for AI-generated responses that cannot authentically fulfill these requirements.

This element aligns closely with authentic assessment principles, which emphasize the importance of connecting academic learning to students' lived experiences and future professional contexts. As Honea et al. (2017) argue, authentic assessment should engage students in applying knowledge to meaningful contexts rather than artificial or decontextualized scenarios. Personalization requirements not only enhance assessment security but also promote this valuable connection between academic content and personal relevance.

Conceptual implementation considerations for personalization include several dimensions. First, experiential integration involves incorporating students' prior experiences or observations into their assessment responses. Second, reflective personalization requires students to consider how course concepts relate to their individual values, goals, or circumstances. Third, applicative personalization demands demonstration of how knowledge might be applied in students' specific future professional contexts. These dimensions provide a basis for operationalizing personalization in assessment design.

However, the theoretical framework must acknowledge potential challenges associated with personalization requirements. Most notably, verification of genuinely personal content presents difficulties, as instructors may lack knowledge of students' actual experiences or circumstances. Additionally, some students might have limited relevant experiences to draw upon or may feel uncomfortable sharing personal information in academic contexts. These challenges highlight the importance of thoughtfully implementing personalization requirements with appropriate flexibility and privacy considerations. Personalization creates a particularly effective barrier against AI-generated content when combined with other elements of the framework. For example, personalization paired with process visibility requirements creates a powerful combination, as it demands not only personal content but also evidence of how that content was integrated into the thinking process. Similarly, personalization combined with temporal relevance might require reflection on recent personal experiences, creating multiple layers of challenge for generative AI.

### 3.5. Resource Accessibility

Resource accessibility emerges as a theoretically significant element in the static analysis framework, focusing on the degree to which assessments require engagement with materials or information that would not be readily available to generative AI systems. This element addresses a key limitation of large language models: their knowledge is constrained to their training data and any additional information explicitly provided in the prompt (Lewis et al., 2020).

The theoretical basis for resource-based constraints stems from the architectural limitations of generative AI systems. Unlike humans, who can access diverse information sources during assessment completion, standard large language models operate with fixed knowledge bases and cannot independently retrieve new information beyond what they were trained on or what is provided in the prompt. By requiring engagement with resources that are not widely available or are accessible only through specific channels, assessments can create barriers for AI-generated responses. This element aligns information literacy frameworks, which emphasize the importance of students developing skills in locating, evaluating, and applying information from diverse sources. As Zawacki-Richter et al. (2019) argue, these skills remain essential even in an AI-augmented educational landscape, as they enable critical engagement with knowledge rather than passive consumption. Resource accessibility requirements not only enhance assessment security but also promote these valuable information literacy skills.

Conceptual measurement approaches for resource accessibility might include evaluating the degree to which assessments require engagement with course-specific materials that are not publicly available, recently created resources that post-date AI training data, or specialized databases and tools that require authenticated access. These requirements create barriers for generative AI, which cannot independently access such resources without explicit provision in the prompt.

However, the theoretical framework must acknowledge potential tensions between resource accessibility requirements and educational equity considerations. Students with different levels of access to specialized resources or those with disabilities affecting their ability to engage with certain resource formats might be disadvantaged by assessments that place excessive emphasis on resource accessibility. These considerations highlight the importance of balancing security concerns with accessibility and inclusiveness in assessment design. Resource accessibility can be conceptualized along several dimensions. First, exclusivity involves the degree to which required resources are restricted to enrolled students rather than publicly available. Second, specificity focuses on how tailored the resources are to the particular course or program rather than general educational materials. Third, format diversity considers whether resources span multiple modalities (text, audio, visual, interactive) that might present different levels of challenge for generative AI. These dimensions provide a theoretical basis for operationalizing resource accessibility in assessment design.

### 3.6. Multimodal Integration

Multimodal integration represents a theoretically distinct element in the static analysis framework, focusing on the degree to which assessments require engagement across different modalities or forms of representation. This element addresses a developing but still limited capability of generative AI: while newer models increasingly incorporate multimodal features, they still face significant challenges in seamlessly integrating across modalities or translating between different representational forms (Ming-Fei et al., 2022).

The justification for multimodal integration as an AI challenge stems from the architectural complexity required for cross-modal reasoning. Even as models like GPT-4V incorporate visual inputs alongside text, they still struggle with tasks requiring deep integration across modalities, such as detailed analysis of visual data, translation between representational forms, or synthesis of information from diverse modal sources (OpenAI, 2023). By requiring such integration, assessments can create barriers for generative AI that excel primarily within rather than across modalities. This element aligns with multiple intelligences theory, which recognizes that human cognition encompasses diverse forms of intelligence beyond verbal-linguistic capabilities (Brookfield, 2017). By incorporating multiple modalities, assessments not only become more resistant to AI-generated responses but also better accommodate diverse learning styles and cognitive strengths. This alignment between security and inclusive pedagogical practice strengthens the case for multimodal integration as an assessment element.

Conceptual implementation considerations for multimodal integration include several dimensions. First, cross-modal analysis involves interpreting information presented in one modality (e.g., visual) and expressing insights in another (e.g., verbal). Second, multimodal synthesis requires integrating information from diverse modal sources to develop comprehensive understanding. Third, representational translation demands conversion between different forms of representation (e.g., translating quantitative data into visual representations or narrative descriptions). These dimensions provide a basis for operationalizing multimodal integration in assessment design.

However, the theoretical framework must acknowledge potential challenges associated with multimodal requirements. Most notably, accessibility concerns arise for students with disabilities affecting their ability to engage with certain modalities. Additionally, technological constraints might limit some students' ability to produce or engage with particular modal forms. These challenges highlight the importance of thoughtfully implementing multimodal integration with appropriate accommodations and alternatives.

Multimodal integration creates particularly effective barriers against AI-generated content when the required modalities extend beyond those commonly integrated in current AI systems. While text-to-image and image-to-text translations are increasingly supported by multimodal AI, other combinations, such as audio analysis, physical artifact creation, or embodied performance, present greater challenges. This observation suggests that the effectiveness of multimodal integration as a security measure will evolve alongside AI capabilities, requiring ongoing refinement of the specific modal combinations employed.

### 3.7. Ethical Reasoning Requirements

Ethical reasoning requirements constitute a theoretically significant element in the static analysis framework, focusing on the degree to which assessments demand nuanced moral judgment, values-based analysis, or engagement with ethical dilemmas. This element addresses a fundamental limitation of generative AI: while large language models can summarize ethical frameworks or simulate ethical reasoning, they lack genuine moral agency or authentic values-based judgment (Brundage et al., 2020). The theoretical basis for ethical reasoning as an AI challenge stems from the nature of moral judgment as a distinctly human capability grounded in consciousness, empathy, and lived experience. Large language models operate through statistical pattern recognition rather than genuine understanding of moral significance or authentic commitment to ethical principles (Floridi & Chiriatti, 2020). By requiring nuanced ethical analysis that goes beyond summarizing established frameworks, assessments can theoretically distinguish between authentic human reasoning and AI-generated content that lacks genuine moral agency.

This element aligns closely with critical thinking frameworks, which emphasize the importance of evaluative judgment and reasoned consideration of complex issues. As Brookfield (2017) argues, developing students' capacity for critical analysis of value-laden situations represents a crucial educational outcome. Ethical reasoning requirements not only enhance assessment security but also promote these valuable critical thinking skills, creating alignment between AI resilience and pedagogical value.

Conceptual measurement approaches for ethical reasoning might include evaluating the degree to which assessments require identification of ethical dimensions in complex situations, application of ethical frameworks to novel contexts, resolution of ethical tensions or dilemmas, or reflection on personal values in relation to professional ethics. These requirements create barriers for generative AI, which can reproduce ethical frameworks but struggles with authentic application to complex, ambiguous situations.

However, the theoretical framework must acknowledge potential tensions between ethical reasoning requirements and other educational considerations. Most notably, assessment of ethical reasoning raises questions about the role of personal values in academic evaluation and the potential for bias in judging students' ethical positions. Additionally, cultural differences in ethical frameworks and reasoning approaches might disadvantage students from backgrounds different from those of the assessors. These considerations highlight the importance of thoughtfully implementing ethical reasoning requirements with sensitivity to diversity and clear evaluation criteria focused on reasoning quality rather than specific conclusions. ethical reasoning can be conceptualized along several dimensions. First, identification involves recognizing ethical dimensions in situations where they might not be immediately apparent. Second, analysis requires application of appropriate ethical frameworks to complex situations. Third, resolution demands working through tensions between competing ethical principles or values. These dimensions provide a theoretical basis for operationalizing ethical reasoning in assessment design while maintaining alignment with educational objectives.

## 3.8. Collaborative Elements

Collaborative elements represent the final component in the static analysis framework, focusing on the degree to which assessments require genuine interaction, negotiation, or co-creation with others. This element addresses a fundamental limitation of generative AI: while large language models can simulate multiple perspectives or voices, they cannot authentically engage in real-time collaboration or genuinely respond to unpredictable human contributions (Kshetri et al., 2024).

The justification for collaboration as an AI challenge stems from the inherently social and interactive nature of genuine collaborative work. Effective collaboration requires responsive adaptation to others' contributions, negotiation of meaning and approach, and integration of diverse perspectives in ways that generative AI cannot authentically replicate. By requiring evidence of genuine collaborative processes, assessments can theoretically distinguish between authentic human interaction and AI-generated content that merely simulates collaboration.

This element aligns closely with social constructivist learning theories, which emphasize that knowledge is constructed through social interaction rather than individual cognition alone. As Carless and Boud (2018) argue, collaborative learning environments enable students to develop crucial skills in negotiation, perspective-taking, and collective problem-solving. Collaborative requirements not only enhance assessment security but also promote these valuable interpersonal skills, creating alignment between AI resilience and pedagogical value.

Conceptual implementation considerations for collaboration include several dimensions. First, interactive collaboration involves real-time engagement with others during the assessment process. Second, integrative collaboration requires synthesizing diverse contributions into coherent outputs. Third, negotiated collaboration demands evidence of how differences in perspective or approach were resolved. These dimensions provide a basis for operationalizing collaboration in assessment design.

However, the framework must acknowledge potential challenges associated with collaborative requirements. Most notably, equity concerns arise regarding group formation, individual contribution assessment, and accommodation of students with social anxiety or other conditions affecting collaborative work. Additionally, logistical challenges in scheduling and coordinating collaborative activities might create barriers for some students, particularly those with significant work or family responsibilities. These challenges highlight the importance of thoughtfully implementing collaborative requirements with appropriate flexibility and support.

collaboration creates particularly effective barriers against AI-generated content when it involves unpredictable human contributions that cannot be anticipated or simulated. While generative AI might produce content that appears collaborative in nature, it cannot authentically respond to the unpredictable, evolving contributions of human collaborators in real time. This observation suggests that the effectiveness of collaboration as a security measure depends on designing requirements that emphasize genuine interaction rather than merely collaborative formats or outputs.

The integration of collaborative elements with other components of the static analysis framework creates powerful combinations for enhancing assessment resilience. For example, collaboration paired with process visibility requirements creates a particularly strong barrier, as it demands evidence of how collaborative thinking evolved over time. Similarly, collaboration combined with ethical reasoning might require negotiation of different value perspectives, creating multiple layers of challenge for generative AI that lacks both genuine moral agency and authentic collaborative capabilities.

## 4. Theoretical Framework for Vulnerability Scoring

### 4.1. Conceptual Basis for Quantitative Assessment

The framework for vulnerability scoring represents a critical component in the machine-versus-machine approach to assessment integrity. This framework provides a conceptual basis for translating qualitative judgments about assessment vulnerability into quantitative measures that enable consistent evaluation, prioritization, and remediation. The justification for numerical vulnerability scoring draws on multiple disciplinary perspectives, including risk assessment theory, psychometrics, and decision science.

From a risk assessment perspective, vulnerability scoring aligns with established frameworks that quantify risk as a function of threat likelihood and potential impact (Brundage et al., 2020). In the context of assessment vulnerability, the likelihood component relates to the probability that generative AI could successfully complete the assessment task, while the impact component concerns the consequences for learning outcomes and academic integrity. By quantifying these dimensions, vulnerability scoring enables more systematic and consistent evaluation than purely qualitative approaches.

The advantages of quantitative assessment over binary classification (vulnerable/not vulnerable) are substantial. First, quantitative scoring acknowledges the continuous rather than dichotomous nature of vulnerability, recognizing that assessments exist on a spectrum of susceptibility to AI-generated responses. This nuanced approach aligns with the reality that few assessments are either completely immune to or entirely vulnerable to generative AI. Second, numerical scoring facilitates comparison across different assessments, enabling educators to prioritize remediation efforts based on relative vulnerability. Third, quantitative measures support tracking of changes over time, allowing for evaluation of whether modifications have successfully reduced vulnerability.

From a psychometric perspective, the framework for vulnerability scoring draws on principles of construct validity and measurement theory. The vulnerability construct is conceptualized as multidimensional, comprising the eight elements of static analysis discussed previously. Each element represents a distinct dimension along which vulnerability can be assessed, with the overall score reflecting a weighted composite of these dimensions. This approach aligns with psychometric principles for measuring complex constructs that cannot be directly observed but must be inferred from multiple indicators (Gregor & Hevner, 2013).

The framework also acknowledges potential challenges in quantitative vulnerability assessment. Most notably, the precision implied by numerical scores might create a false sense of objectivity or certainty about what remains an inherently probabilistic judgment. Additionally, the complexity of the vulnerability construct raises questions about whether a single composite score can adequately capture the multidimensional nature of assessment resilience. These challenges highlight the importance of transparent scoring methodologies and appropriate communication of confidence levels in vulnerability assessments.

## 4.2. Weighting Theoretical Framework

The theoretical framework for differential weighting of vulnerability elements addresses the recognition that not all dimensions contribute equally to overall assessment resilience. This framework provides a conceptual basis for determining the relative importance of different elements in the vulnerability scoring system, ensuring that the composite measure accurately reflects the understanding of what makes assessments susceptible to generative AI.

The conceptual basis for differential weighting aligns with assessment validity principles, particularly the notion that validity evidence should be weighted according to its relevance to the specific inference being made (Biggs & Tang, 2022). In the context of vulnerability scoring, this principle suggests that elements more directly predictive of AI performance on assessment tasks should receive greater weight than those with more indirect or contextual relationships to vulnerability.

Weighing can be approached through several methodologies, each with distinct conceptual foundations. First, empirical weighting would derive weights from observed relationships between element scores and actual AI performance across diverse assessment types. While this approach offers data-driven precision, it faces limitations related to the rapidly evolving nature of AI capabilities, potentially rendering historically derived weights obsolete as new models emerge.

Second, expert judgment weighting would rely on consensus among assessment specialists, AI researchers, and educational theorists regarding the relative importance of different elements. This approach draws on the value of collective expertise in domains characterized by complexity and uncertainty. The Delphi method, involving structured iterative feedback from experts, provides a theoretical framework for developing consensus weights while minimizing individual biases or limited perspectives (Hevner et al., 2004).

Third, derivation would establish weight based on conceptual analysis of how each element relates to fundamental limitations in generative AI architecture and capabilities. This approach aligns with the understanding that certain constraints (such as training data cutoffs or limited multimodal integration) represent more persistent limitations than others that might be overcome through technical advancement. Weights derived from analysis might prove more stable over time than those based solely on current AI performance.

The theoretical framework must also address potential tensions in weighing across different assessment contexts. Elements that prove highly predictive of vulnerability in one discipline or assessment type might be less relevant in others. For example, temporal relevance might be particularly important in rapidly evolving fields like technology or current affairs but less critical in disciplines focused on historical analysis or foundational principles. This contextual variation suggests the value of discipline-specific weighting schemes that reflect the particular characteristics and priorities of different educational domains.

Additionally, the weighting framework must consider how interactions between elements might influence overall vulnerability. Certain combinations of elements might create synergistic effects that exceed the sum of their individual contributions. For instance, the combination of high specificity and strong process visibility requirements might create a particularly effective barrier against AI-generated responses, suggesting the importance of considering interaction effects in the weighting system.

### 4.3. Threshold Determination Theory

The theoretical framework for vulnerability thresholds addresses the need to translate continuous vulnerability scores into actionable categories that guide educator decision-making. This framework provides a conceptual basis for determining at what points along the vulnerability spectrum assessments should be classified as requiring different levels of attention or intervention.

The conceptual basis for threshold determination draws on decision theory, particularly the notion of action thresholds that trigger specific responses when certain conditions are met (Peffers et al., 2007). In the context of vulnerability scoring, these thresholds define the boundaries between different risk categories, each associated with distinct recommendations for assessment modification or monitoring.

The traffic-light categorization (red, amber, green) offers several advantages over more complex classification schemes. First, it provides intuitive understanding through familiar symbolism, reducing cognitive load for educators interpreting vulnerability assessments. Second, it creates clear decision boundaries while maintaining sufficient nuance to distinguish between different levels of concern. Third, it aligns with established risk communication practices across multiple domains, facilitating integration with broader institutional risk management frameworks.

The justification for specific threshold values involves balancing several considerations. First, thresholds must reflect meaningful differences in vulnerability that warrant distinct responses. Setting thresholds too close together might create artificial distinctions without practical significance, while thresholds too far apart might fail to capture important variations in vulnerability. Second, thresholds should consider the resource implications of different classifications, recognizing that institutional capacity for assessment redesign is finite. Third, thresholds must acknowledge the inherent uncertainty in vulnerability prediction, potentially incorporating confidence intervals or tolerance ranges rather than precise cutoff points.

threshold determination can be approached through several methodologies. Empirical approaches would establish thresholds based on observed relationships between vulnerability scores and actual AI performance, identifying natural breakpoints that distinguish between different levels of susceptibility. Normative approaches would set thresholds based on institutional or disciplinary standards regarding acceptable levels of vulnerability, reflecting value judgments about appropriate risk tolerance in educational assessment. Pragmatic approaches would establish thresholds based on practical considerations such as resource availability for remediation, balancing ideal security with realistic implementation constraints.

The theoretical framework must also address how thresholds might vary across different educational contexts. Factors such as academic level (undergraduate vs. graduate), assessment stakes (formative vs. summative), and disciplinary norms might all influence appropriate threshold values. This contextual variation suggests the value of flexible threshold frameworks that can be calibrated to specific institutional or programmatic needs while maintaining conceptual consistency in the overall approach.

Additionally, the threshold framework must consider temporal dynamics in threshold determination. As generative AI capabilities evolve, thresholds that are appropriately distinguished between vulnerability levels at one point might require recalibration to remain meaningful. This temporal dimension suggests the importance of regular threshold review and adjustment, ensuring that vulnerability classifications remain aligned with current technological realities rather than historical benchmarks.

To visualize assessment vulnerability across multiple dimensions, a radar chart can be used to represent scores assigned to each of the eight static analysis elements. This graphical format provides a holistic view of where an assessment is most susceptible to generative AI exploitation and highlights areas of relative strength. By plotting these scores on a shared scale, educators can quickly identify imbalances, prioritize redesigning efforts, and track improvements over time. An example radar chart illustrating a hypothetical assessment's vulnerability profile is shown below.

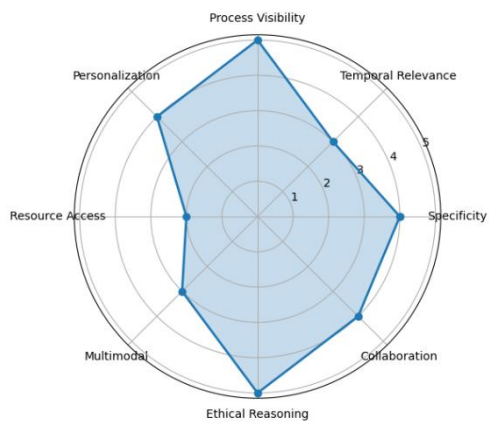

Figure 2 - Assessment Vulnerability Radar Chart Sample

## 5. Implications and Future Directions

### 5.1. Implications for Assessment Design

The dual strategy approach to assessment vulnerability analysis carries profound implications for assessment design in higher education. By integrating static analysis with dynamic testing, this framework fundamentally reconceptualizes how we understand and address the challenges posed by generative AI. These implications extend beyond immediate practical applications to influence broader assessment theory, educational philosophy, and pedagogical practice.

At the most fundamental level, the dual strategy approach challenges the traditional dichotomy between assessment security and pedagogical value. Conventional responses to academic integrity threats often create tension between security measures (such as proctoring or detection) and educational quality. In contrast, the framework presented in this paper demonstrates how elements that enhance resilience to generative AI can simultaneously strengthen pedagogical effectiveness. Specificity and contextualization not only reduce vulnerability but also enhance alignment with specific learning outcomes. Process visibility requirements not only challenge AI simulation but also develop valuable metacognitive skills. This alignment suggests that addressing generative AI challenges can catalyze positive transformations in assessment practice rather than merely defensive reactions.

The framework also has significant implications for assessment theory, particularly regarding the concept of validity in an AI-augmented educational landscape. Traditional validity frameworks focus on whether assessments accurately measure intended learning outcomes (Biggs & Tang, 2022). The dual strategy approach extends this concept to consider whether assessments can distinguish between authentic human learning and AI-generated simulation. This extension suggests a reconceptualization of validity that incorporates resilience to technological mediation as a core component rather than an external consideration. Such reconceptualization may become increasingly important as the boundaries between human and AI-augmented performance continue to blur.

The dual strategy approach also challenges conventional thinking about the relationship between standardization and personalization in assessment. Traditional assessment paradigms often prioritize standardization to ensure comparability and fairness. However, the framework presented in this paper suggests that increased personalization, through elements such as unique contexts, individual experiences, and collaborative interactions, may actually enhance assessment integrity in an AI-augmented environment. This tension between standardization and personalization represents a fundamental challenge to assessment design that requires careful navigation and ongoing development.

The framework further implies a shift in understanding of assessment as a static artifact to assessment as an evolving process. The rapidly advancing capabilities of generative AI mean that assessment vulnerability is not a fixed characteristic but a dynamic relationship between assessment design and technological capability. This temporal dimension suggests the importance of designing assessments with adaptability and evolution in mind, potentially incorporating mechanisms for ongoing refinement and adjustment as AI capabilities advance. Such a process-oriented conceptualization represents a significant departure from traditional views of assessment as a stable, reusable product.

Finally, the dual strategy approach has implications for the role of technology in assessment design. Rather than positioning technology either as a threat to be mitigated or a tool to be leveraged, this framework suggests a more nuanced perspective in which technology simultaneously creates challenges and opportunities. By using AI to analyze and enhance assessment resilience to AI, the approach embodies a reflexive relationship with technology that acknowledges both its limitations and its potential contributions to educational practice.

This nuanced perspective may offer valuable insights for navigating other technological disruptions in education beyond generative AI.

## 5.2. Limitations and Challenges

Despite its significant contributions, the dual strategy approach to assessment vulnerability analysis faces several limitations and challenges that must be acknowledged. These limitations do not undermine the framework's value but highlight areas requiring further development and refinement.

A fundamental limitation concerns the inherent unpredictability of generative AI advancement. The framework assumes that certain patterns of vulnerability can be identified and addressed through systematic analysis. However, the rapid and sometimes unexpected evolution of AI capabilities means that today's understanding of vulnerability may quickly become outdated. This limitation creates a tension between the need for stable assessment principles and the reality of continuously evolving technological capabilities. Addressing this tension requires ongoing work on how assessment frameworks can maintain conceptual stability while accommodating technological change.

The framework also faces challenges related to the balance between security and accessibility. Elements that enhance resilience to generative AI, such as multimodal integration, collaborative requirements, or resource constraints, may simultaneously create barriers for students with disabilities, limited technological access, or specific learning needs. This tension highlights a dilemma in assessment design: how to enhance security without compromising inclusivity. Resolving this dilemma requires deeper integration between assessment security frameworks and universal design principles, ensuring that resilience to AI does not come at the expense of educational equity.

Another limitation concerns the potential for strategic adaptation by users of generative AI. The framework assumes that certain assessment characteristics create barriers for AI-generated responses. However, as students become more sophisticated in their use of these tools, employing techniques such as prompt engineering, model fine-tuning, or hybrid human-AI approaches, the barriers may prove less effective than anticipated. This adaptive dynamic creates a challenge similar to that faced in cybersecurity: how to develop robust defenses against adversaries who continuously evolve their strategies. Addressing this challenge requires theoretical work on anticipating and modeling strategic adaptation rather than assuming static vulnerability patterns.

The framework also faces critiques regarding its potential to reinforce traditional conceptions of assessment that may not align with emerging educational paradigms. By focusing on distinguishing between human and AI-generated work, the approach might implicitly reinforce the primacy of individual cognitive production over collaborative knowledge construction or technological augmentation. This tension highlights a deeper question about the purpose of assessment in an increasingly AI-augmented world: should we assess what students can do without technological assistance, or should we evaluate how effectively they can leverage available tools? Resolving this question requires engagement with fundamental philosophical perspectives on the relationship between technology and human cognition.

Additionally, the dual strategy approach faces challenges related to cultural and disciplinary variation in assessment practices. The framework's elements may have different relevance or manifestation across diverse educational contexts, raising questions about its universal applicability. For example, process visibility requirements might align well with Western educational traditions that emphasize explicit articulation of thinking but conflict with approaches that value implicit understanding or holistic demonstration. Similarly, the relative importance of different elements may vary substantially across disciplines, with temporal relevance perhaps more critical in rapidly evolving fields than in those focused on historical or foundational knowledge. These variations highlight the need for cultural and disciplinary sensitivity in applying the framework, potentially requiring adaptation rather than uniform implementation.

Finally, the framework faces limitations related to the balance between comprehensive analysis and practical implementation. The eight elements of static analysis, combined with dynamic testing, create a theoretically robust but potentially complex approach to vulnerability assessment. This complexity raises questions about whether the framework can be effectively operationalized without simplification that might compromise its integrity. Addressing this limitation requires work on how complex frameworks can be translated into accessible implementation guidance without losing their conceptual richness.

### 5.3. Future Directions

The framework presented in this paper opens numerous avenues for future research and conceptual development. These directions extend beyond refinement of the current approach to explore broader implications for assessment theory, educational philosophy, and the evolving relationship between human and artificial intelligence in educational contexts.

A primary direction for future work involves deeper integration with learning analytics and educational data science. The dual strategy approach generates rich data about assessment characteristics and vulnerability patterns that could inform more sophisticated models of the relationship between assessment design and student learning outcomes. Future research might explore how vulnerability analysis could be combined with learning analytics to develop models that predict not only AI performance on assessments but also patterns of student engagement, learning, and development. Such integration could advance understanding of how assessment design influences both security and pedagogical effectiveness.

Another promising direction involves exploration of adaptive assessment possibilities. The current framework focuses on analyzing and enhancing existing assessments, but future work might consider how the principles could inform dynamically generated assessments that adapt to evolving AI capabilities. Theoretical models of adaptive assessment could explore how elements might be automatically recalibrated based on detected vulnerability patterns or how personalization might be dynamically increased in response to emerging AI capabilities. This direction connects assessment security with broader theoretical work on adaptive learning systems and personalized education.

Future development should also address cross-disciplinary applications and variations. While the current framework provides a general approach to assessment vulnerability, different disciplines likely require specific adaptations. Future research might explore how the eight elements manifest in disciplines with distinct epistemological traditions, methodological approaches, and assessment practices. For example, how might process visibility requirements differ between humanities, sciences, and professional fields? How might temporal relevance operate differently in historical versus contemporary studies? These disciplinary variations require theoretical elaboration to ensure the framework's relevance across diverse educational contexts.

Policy implications represent another critical direction for future work. As institutions develop responses to generative AI, they require theoretical frameworks that can inform coherent, principled approaches rather than reactive policies. Future research might explore how the dual strategy approach could inform institutional policy development, regulatory frameworks, and quality assurance processes. This direction would connect assessment vulnerability analysis with broader theoretical work on educational governance and policy implementation in technological contexts.

The philosophical implications of the dual strategy approach also merit further exploration. By focusing on distinguishing between human and AI-generated work, the framework raises fundamental questions about what we value in education and how we conceptualize authentic human learning. Future philosophical work might examine the ontological and epistemological assumptions underlying the framework, exploring alternative conceptions of the relationship between human cognition, technological augmentation, and educational assessment. This direction would connect assessment security with deeper philosophical questions about technology, humanity, and knowledge in contemporary society.

Longitudinal models represent another important direction for future work. The current framework provides a snapshot approach to vulnerability assessment, but the rapidly evolving nature of both AI capabilities and educational practices suggests the need for models that account for temporal dynamics. Future research might develop frameworks for tracking vulnerability patterns over time, predicting emerging challenges, and modeling the co-evolution of assessment design and AI capabilities. Such longitudinal perspectives would enhance the framework's adaptability to technological change.

Finally, future work should explore the broader implications of the machine-versus-machine paradigm beyond assessment security. The approach of using AI to analyze and enhance resilience to AI might have applications in other educational contexts, such as curriculum design, pedagogical practice, or educational resource development. By extending the principles to these domains, future research could develop a more comprehensive understanding of how educational systems might adapt to and thrive within an increasingly AI-augmented landscape.

## 6. Conclusion

The emergence of generative artificial intelligence has fundamentally altered the landscape of higher education, presenting both unprecedented challenges and remarkable opportunities for assessment practices. This paper has explored a theoretical framework for a machine-versus-machine approach to addressing these challenges, demonstrating how a dual strategy combining static analysis and dynamic testing can provide a comprehensive foundation for understanding and enhancing assessment resilience to generative AI.

The contributions of this research span multiple domains. First, we have established a conceptual framework for understanding how automated vulnerability analysis can identify assessment weaknesses that manual review processes might miss. This framework integrates principles from cybersecurity, educational theory, and assessment design, creating a robust foundation for future work in this area. The conceptualization of assessment vulnerability as a multidimensional construct, encompassing eight distinct elements from specificity to collaborative requirements, provides a valuable theoretical lens for analyzing and improving assessment design in an AI-augmented educational landscape.

Second, we have provided a detailed justification for the dual strategy approach, explaining how the complementary strengths of static analysis and dynamic testing can overcome their respective limitations. Static analysis offers scalable, pattern-based evaluation but may miss emerging or context-specific vulnerabilities; dynamic testing captures specific performance but may not identify underlying patterns. By integrating these approaches, the dual strategy paradigm offers both breadth and depth in vulnerability assessment, enabling both program-level coverage and detailed insight into particular assessments.

Third, we have presented a comprehensive justification for each of the eight elements of static analysis. Specificity and contextualization address the limitations of AI in generating responses to unique or highly specialized prompts. Temporal relevance leverages the training data cutoff constraints of large language models. Process visibility requirements exploit the difference between human cognitive development and AI token prediction. Personalization elements challenge AI's inability to authentically incorporate individual experiences. Resource accessibility, multimodal integration, ethical reasoning, and collaborative elements each address specific limitations in current generative AI capabilities.

Fourth, we have developed a framework for vulnerability scoring that translates qualitative judgments into quantitative measures, enabling consistent evaluation and prioritization. This framework addresses the conceptual basis for numerical assessment, the justification for differential weighting of elements, and the decision theory underlying threshold determination. By providing a foundation for these components, the paper offers a comprehensive approach to operationalizing vulnerability assessment while maintaining conceptual rigor.

The broader implications of this research extend beyond the specific context of assessment vulnerability. The dual strategy paradigm challenges traditional dichotomies between assessment security and pedagogical value, demonstrating how elements that enhance resilience to generative AI can simultaneously strengthen educational effectiveness. It extends conventional validity frameworks to consider resilience to technological mediation as a core component rather than an external consideration. It challenges traditional thinking about standardization and personalization, suggesting that increased personalization may actually enhance assessment integrity in an AI-augmented environment.

Despite these contributions, several limitations and challenges must be acknowledged. The inherent unpredictability of generative AI advancement creates tension between the need for stable assessment principles and the reality of continuously evolving technological capabilities. The balance between security and accessibility raises dilemmas about enhancing resilience without compromising inclusivity. The potential for strategic adaptation by users of generative AI creates challenges similar to those faced in cybersecurity. Cultural and disciplinary variations in assessment practices raise questions about the universal applicability of the framework.

Future directions include deeper integration with learning analytics, exploration of adaptive assessment possibilities, elaboration of cross-disciplinary applications, development of policy implications, examination of philosophical assumptions, creation of longitudinal models, and extension of the machine-versus-machine paradigm beyond assessment security. By pursuing these directions while addressing current limitations, researchers and educators can continue to develop effective responses to generative AI challenges that enhance rather than compromise educational quality and integrity.

In conclusion, the framework presented in this paper offers a promising foundation for understanding and addressing the challenges posed by generative AI in higher education assessment. By leveraging a dual strategy approach that combines static analysis with dynamic testing, educators can develop more nuanced and effective approaches to maintaining assessment integrity in an era of rapidly evolving technological capabilities. Rather than viewing generative AI solely as a threat to be mitigated, this framework recognizes its potential to catalyze positive transformations in assessment practice, ultimately enhancing the quality and relevance of higher education in a technologically augmented future.